\documentclass[longauth]{aa}  

\usepackage{graphicx}
\usepackage{txfonts}
\usepackage{threeparttable}
\usepackage{hyperref}
\hypersetup{colorlinks = true,linkcolor= blue,citecolor=blue}
\usepackage{chemformula}

\long\def\ce#1{\ensuremath{#1}}

\begin{document} 

\title{JWST observations of \ce{^{13}CO_2} ice:}
\subtitle{Tracing the chemical environment and thermal history of ices in protostellar envelopes}

\author{Nashanty G. C. Brunken\inst{1}, Will R. M. Rocha\inst{1,3}, Ewine F. van Dishoeck \inst{1,4}, Robert Gutermuth\inst{4}, Himanshu Tyagi\inst{5}, Katerina Slavicinska\inst{1,3}, Pooneh Nazari\inst{1}, S. Thomas Megeath\inst{2}, Neal J. Evans II\inst{9}, Mayank Narang \inst{6}, P. Manoj\inst{5}, Adam E. Rubinstein\inst{7}, Dan M. Watson\inst{8},  Leslie W. Looney\inst{10,11}, Harold Linnartz\inst{1,3}, Alessio Caratti o Garatti\inst{12}, Henrik Beuther\inst{13}, Hendrik Linz\inst{13,14}, Pamela Klaassen\inst{15}, Charles A. Poteet\inst{29}, Samuel Federman\inst{2}, Guillem Anglada\inst{17}, Prabhani Atnagulov\inst{2}, Tyler L. Bourke\inst{18}, William J. Fischer\inst{16}, Elise Furlan\inst{19}, Joel Green\inst{16}, Nolan Habel\inst{20}, Lee Hartmann\inst{21}, Nicole Karnath\inst{22,23}, Mayra Osorio\inst{17}, James Muzerolle Page\inst{16}, Riwaj Pokhrel\inst{2},Rohan Rahatgaonkar\inst{24}, Patrick Sheehan\inst{25}, Thomas Stanke\inst{4}, Amelia M.\ Stutz\inst{26}, John J. Tobin\inst{11}, Lukasz Tychoniec\inst{27}, Scott Wolk\inst{23} and Yao-Lun Yang\inst{28} }

\institute{Leiden Observatory, Leiden University, 2300 RA Leiden, The Netherlands \\
\email{brunken@strw.leidenuniv.nl} 
\and
Department of Physics and Astronomy, The University of Toledo, 2801 West Bancroft Street, Toledo, OH 43606, USA 
\and
Laboratory for Astrophysics, Leiden Observatory, Leiden University, PO Box 9513, 2300 RA Leiden, The Netherlands 
\and 
Max-Planck-Institut f\"{u}r Extraterrestrische Physik, Gie{\ss}enbachstra{\ss}e 1, 85748 Garching, Germany 
\and
University of Massachusetts Amherst, Amherst, MA, USA 
\and 
Tata Institute of Fundamental Research, Mumbai 400005, India 
\and
Academia Sinica Institute of Astronomy \& Astrophysics, 11F of Astro-Math Bldg., No.1, Sec. 4, Roosevelt Rd., Taipei, Taiwan 
\and
Department of Physics and Astronomy, University of Rochester, 500 Wilson Boulevard, Rochester, NY, 14611, USA 
\and
Department of Astronomy, The University of Texas at Austin, 2515 Speedway, Stop C1400, Austin, Texas 78712-1205, USA 
\and
Department of Astronomy, University of Illinois, 1002 W. Green St., Urbana, IL, 61801 
\and
National Radio Astronomy Observatory, 520 Edgemont Rd., Charlottesville, VA 22903, USA 
\and
INAF-Osservatorio Astronomico di Capodimonte, Italy 
\and
Max Planck Institute for Astronomy, Heidelberg, Baden Wuerttemberg, Germany 
\and
Friedrich-Schiller-Universit\"at, Jena, Th\"uringen, Germany 
\and
United Kingdom Astronomy Technology Centre, Edinburgh, United Kingdom 
\and
Space Telescope Science Institute, 3700 San Martin Drive, Baltimore, MD 21218, USA 
\and
Instituto de Astrofísica de Andalucía, CSIC, Glorieta de la Astronomía
s/n, E-18008 Granada, Spain 
\and
SKA Observatory, Jodrell Bank, Lower Withington, Macclesfield SK11 9FT, United Kingdom 
\and
Caltech/IPAC, Pasadena, CA, USA 
\and
Jet Propulsion Laboratory, Pasadena, USA 
\and
University of Michigan, Ann Arbor, MI, USA 
\and
Space Science Institute, Boulder, CO, USA 
\and
Center for Astrophysics Harvard \& Smithsonian, Cambridge, MA, USA 
\and
Gemini South Observatory, La Serena, Chile 
\and
Northwestern University, Evanston, Illinois, USA 
\and
Departamento de Astronom\'{i}a, Universidad de Concepci\'{o}n,Casilla 160-C, Concepci\'{o}n, Chile 
\and
European Southern Observatory,
Garching bei M\"{u}nchen, Germany 
\and
RIKEN Cluster for Pioneering Research, Wako-shi, Saitama, 351-0106, Japan 
\and
NV5 Geospatial Solutions, Inc. 385 Interlocken Crescent, Suite 300 Broomfield, CO 80021, USA 
}

\titlerunning{}
\authorrunning{N.G.C. Brunken et al.}

\date{Received; Accepted}

\abstract{The structure and composition of  simple ices can be severely modified during stellar evolution by protostellar heating. Key to understanding the involved processes are thermal and chemical tracers that can be used to diagnose the history and environment of the ice. The 15.2 \ce{\mu}m bending mode of \ce{^{12}CO_2} in particular has proven to be a valuable tracer of ice heating events but suffers from grain shape and size effects. A viable alternative tracer is the weaker \ce{^{13}CO_2} isotopologue band at 4.39 \ce{\mu}m, which has now become accessible at high S/N with the \textit{James Webb} Space Telescope (JWST). In this study we present JWST NIRSpec observations of \ce{^{13}CO_2 } ice in five deeply embedded Class 0 sources that span a wide range in masses and luminosities  (0.2 - \ce{10^4} $L_\odot$ ) taken as part of the Investigating Protostellar Accretion Across the Mass Spectrum (IPA) program. The band profiles vary significantly depending on the source, with the most luminous sources showing a distinct narrow peak at 4.38 \ce{\mu}m. We first applied a phenomenological approach with which we demonstrate that a minimum of three to four Gaussian profiles are needed to fit the absorption feature of \ce{^{13}CO_2}. We then combined these findings with laboratory data and show that a 15.2 \ce{\mu}m \ce{^{12}CO_2} bending-mode-inspired five-component decomposition can be applied to the isotopologue band, with each component representative of \ce{CO_2} ice in a specific molecular environment. The final solution consists of cold mixtures of \ce{CO_2} with \ce{CH_3OH}, \ce{H_2O}, and CO as well as segregated heated pure \ce{CO_2} ice at 80 K. Our results are in agreement with previous studies of the \ce{^{12}CO_2} ice band, further confirming that \ce{^{13}CO_2} is a useful alternative tracer of protostellar heating and ice composition. We also propose an alternative solution consisting only of heated mixtures of \ce{CO_2}:\ce{CH_3OH} and \ce{CO_2}:\ce{H_2O} ices and warm pure \ce{CO_2} ice at 80 K (i.e., no cold \ce{CO_2} ices) for decomposing the ice profiles of HOPS 370 and IRAS 20126, the two most luminous sources in our sample that show strong evidence of ice heating resulting in ice segregation.}


\maketitle


\section{Introduction} 

Simple interstellar ices have long been the focus of many studies because they offer insight into the processes that take place on icy grains. These processes are expected to ultimately result in the formation of complex organic molecules, the ingredients required for habitability \citep{herbst2009complex, mumma2011chemical, boogert2015observations}. Their abundance and composition set the initial conditions of the parent molecular cloud from which new stars and later planets are formed. These initial conditions, however, can be severely altered by the violent processes that take place during stellar evolution, when ices can be physically and chemically modified by thermal processing and, in extreme cases, even destroyed \citep{visser2009chemical,oberg2011spitzer,  mumma2011chemical, brownlee2014stardust}. Moreover, changes in the ice composition and structure can serve as unique tracers, most notably of ice heating due to, for example, protostellar accretion bursts. Knowledge on how ices mature after their formation is therefore essential if we are to understand how they are incorporated into planetary material. Ground- and space-based observatories have provided a wealth of information on these simple ices through infrared spectroscopy \citep[e.g.,][]{gibb2000inventory,boogert2002high,pontoppidan2003m,gibb2004interstellar, boogert2008c2d, boogert2015observations, oberg2011spitzer,yang2022corinos, mcclure2023ice}. 

In the solid phase, molecules can vibrate and absorb infrared light at particular wavelengths, thereby producing infrared ice absorption bands. When studied in combination with laboratory data, useful information about the ice properties can be extracted from these absorption bands \citep{tielens1991interstellar,allamandola1999evolution,boogert2015observations}. Carbon dioxide (\ce{CO_2}) in particular has proven to be a valuable tracer for diagnosing the chemical composition and thermal history of interstellar ices. First hints of \ce{CO_2} ice emerged when its 15.2 \ce{\mu}m band was \text{observed} at low resolution with the Infrared Astronomical Satellite (IRAS) satellite \citep{d1989discovery}, and a firm detection was later confirmed by \citet{graauw1996sws} using observations from the Infrared Space Observatory (ISO). 
Generally considered to be the second most abundant ice species after water \citep[\ce{H_2O};][]{ioppolo2011surface, oberg2011spitzer}, \ce{CO_2} has been observed in multiple environments \citep{graauw1996sws,whittet1998detection, gerakines1999infrared,bergin2005spitzer, whittet2007abundance,whittet2009nature, mcclure2023ice}. Moreover, in the laboratory, accurate spectroscopic data have been recorded for \ce{CO_2} ice under a number of conditions, and several solid-state pathways toward \ce{CO_2} have been characterized \citep{palumbo1999solid, oberg2007effect, isokoski2013highly, ioppolo2013surfre}, thus making  \ce{CO_2} ideal for studying ices across different astronomical sources. 

A \textit{Spitzer} Space Telescope survey conducted by \citet{pontoppidan2008c2d} investigated the intriguing variations seen in the 15.2 \ce{\mu}m bending mode of \ce{^{12}CO_2}, including at times a characteristic double peak structure, when looking at different lines of sight. These variations were also observed in experimental results \citep{ehrenfreund1997infrared} and during other studies with ISO \citep{gerakines1999infrared} and \textit{Spitzer} \citep{boogert2004spitzer,knez2005spitzer, zasowski2009spitzer, kim2012co2, poteet2013anomalous}. \citet{pontoppidan2008c2d} analyzed this band for 50 low-mass protostars and established a method for decomposing the ice profile of each source using five unique components, each individual component corresponding to \ce{CO_2} in a specific molecular ice environment. Consistent with previous analyses of this ice band, they found that the \ce{^{12}CO_2} bending mode comprises polar water-rich ices and apolar CO-rich ices.

Despite being a sensitive tracer, the 15.2 \ce{\mu}m band does have one main disadvantage. Similar to all the strong absorption bands of the dominant ices, the bending mode of \ce{CO_2} is highly susceptible to grain shape and size effects \citep{ehrenfreund1996laboratory,baratta2000,  dartois2006spectroscopic, dartois2022influence}. These effects can distort the regular appearance of the ice absorption bands and consequently influence the spectral analysis. As a result, optical constants need to be calculated and included in the band analysis to account for these modifications. These grain shape and size effects become much smaller or even negligible, however, for the weaker ice bands \citep{tielens1991interstellar, ehrenfreund1997infrared}. Absorption bands of isotopologues in particular, which are usually expected to be substantially less abundant than their main counterparts, therefore offer viable alternatives. 

Experimental studies by \citet{ehrenfreund1997infrared} revealed that the asymmetric stretching mode of \ce{^{13}CO_2} at 4.39 \ce{\mu}m also displays strong peak shifts and profile broadening when the conditions of the ice matrix are altered. These laboratory data were later used to aid in the interpretation of ISO-SWS observations of high-mass protostars, in which the observed profiles of the \ce{^{13}CO_2} ice bands vary significantly depending on the source \citep{boogert1999iso}. The main finding was a blue peak that appears in a number of high-mass sources. This was attributed to either pure \ce{CO_2} in heated polar ices containing diluted amounts of methanol (\ce{CH_3OH}) ice or heated \ce{CO_2}:\ce{H_2O}:\ce{CH_3OH} ices. These results further confirmed the significance of this weaker ice band and how it can be used to diagnose interstellar ice morphologies. 

Now with the high sensitivity and spectral resolution of the \textit{James Webb} Space Telescope (JWST) and high resolution laboratory data, we are well equipped to study these weak ice features in much greater detail. In this paper we present a spectral analysis of the \ce{^{13}CO_2} isotopologue band for five Class 0 embedded sources observed with the Near Infrared Spectrograph (NIRSpec) on board the JWST as part of the Investigating Protostellar Accretion Across the Mass Spectrum (IPA) program \citep{federman2023,narang2023,rubinstein2023}. The wide luminosity range of the sample (0.2 - \ce{10^4} $L_\odot $) allows us to study the effect of protostellar ice heating for low- and high-mass sources. Furthermore, the exceptional sensitivity of the JWST will make it possible to study this weak band for low-mass protostars at very high signal/noise. 

We intend to build on the works of \citet{boogert1999iso} and \citet{pontoppidan2008c2d} by methodologically decomposing the isotopologue band and investigating to what extent its components are similar or different to those of the 15.2 \ce{\mu}m bending mode of \ce{^{12}CO_2}. Additionally, this study includes laboratory data taken at a higher spectral resolution \citep[0.5 \ce{cm^{-1}};][]{van2006infrared}. In Section \ref{sec:2} we present our observations and describe the methods used to analyze the bands. In Section \ref{sec:3} we present the spectral decomposition, and the results are discussed in Section \ref{sec:4}. Finally, in Section \ref{sec:5} we summarize the main points of this study and provide our conclusions.

\section{Data and methods}
\label{sec:2}

\begin{center}
\begin{table*}[hbt!]
\caption{Properties of the sources observed in this sample.}
\small
\centering
\begin{tabular}{lcccc}
\hline \hline
Source & Distance (pc) & Luminosity ($L_\odot $ )  & Stellar Mass ($M_\odot$ ) & References
\\     
\hline

IRAS 16253 & 140 & 0.16 & 0.12 - 0.17 & 5, 1, 1 \\
B335 & 165 & 1.4 & 0.25 & 6, 2, 2  \\
HOPS 153 & 390 & 3.8 & 0.6 & 7, 3, 9 \\
HOPS 370 & 390 & 310 & 2.5 & 7, 3, 10  \\
IRAS 20126 & 1550 & \ce{10^4} & 12 & 8, 4, 4  \\

\hline
\end{tabular}
\label{Tab:1}
\begin{tablenotes}\footnotesize
\item{\textbf{Notes.} References: 1: \citet{aso2023early}, 2: \citet{evans2023models}, 3: \citet{furlan2016herschel}, 4: \citet{chen2016hot}, 5: \citet{ortiz2018gaia}, 6: \citet{watson2020distance}, 7: \citet{tobin2022vla}, 8: \citet{reid2019trigonometric} 9: J. Tobin priv. comm., 10: \citet{tobin2020vla} } 
\end{tablenotes}
\end{table*}  
\end{center}

\begin{figure*}[h!]
    \centering
     \includegraphics[width=1\hsize]{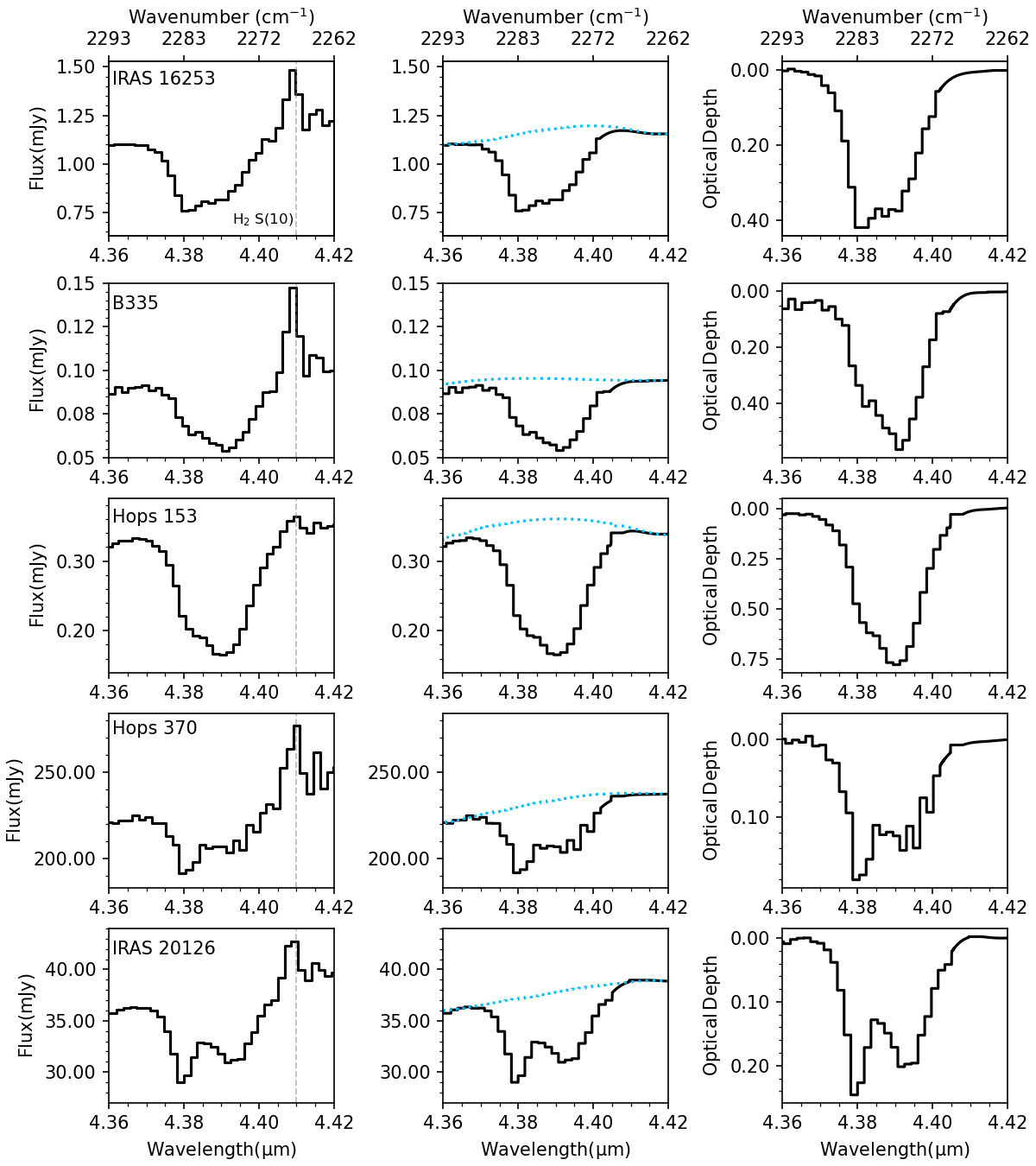}
    \caption{Overview of the \ce{^{13}CO_2} ice feature toward the IPA sources studied in this work in order of increasing source luminosity. \textit{First column:} Data before subtracting the \ce{H_2} and the CO ro-vibrational lines and the continuum. The dashed gray line shows the position of the \ce{H_2} 0 - 0 S(10) emission line. \textit{Second column:} Bands after subtracting CO gas-phase lines. The dotted blue line shows the local continuum that was traced for each band. \textit{Third column:} Bands in optical depth scale after continuum subtraction.} 
    \label{fig:overview}
\end{figure*}

\subsection{Observations}

The sample consists of five class 0 protostars observed as part of the IPA Cycle 1 GO program (PI: T.Megeath, ID: 1802). An overview of the properties of each source is provided in Table \ref{Tab:1}. The sample was observed using the G395M mode \citep[R = $\lambda$ /$\Delta$ $\lambda$ = 1250;][]{rubinstein2023} of the NIRSpec integral field unit (IFU) in a 2 x 2 mosaic with 10\% overlap, using the four-point dither mode with a spatial resolution of 0.2'', corresponding to 30 to 300 AU. The distances of the protostars in our sample vary from 140 to 1550 pc. The data reduction process is covered in detail in \citet{federman2023}. 

For the purpose of the ice analysis, we extracted spectra from the image cubes at a central position at the source continuum using an aperture size of 0.6'' to ensure no flux was excluded. The extraction coordinates for each sources are given in Table \ref{Tab:coordinates}. We estimated the effect of the radial velocity for the sources in our sample and found that even for the most massive source IRAS 20126, the effect is negligible given that the radial velocity is rather small (\ce{V_{lsr}} = -3.5 \ce{km \, s^{-1}}). The same holds for the other sources in this sample and given our data precision we expect that any Doppler shifts due to radial velocities are small and hidden in the noise during the wavelength calibration.

\subsection{Gas-phase lines and continuum removal}

In order to properly determine the continuum it was necessary to first isolate the ice bands from the CO ro-vibrational gas-phase lines that are prominent in the spectra of all five sources as shown in Figures \ref{fig:Hops370-isolated} and \ref{fig:Hops370-hot}. The gas-phase lines are further discussed in \citet{rubinstein2023}. Our method consists in tracing the gas-phase lines and selecting points at the bottom of the traced gas-phase lines to fit a spline function. The fitted spline function then serves as a continuum and we added the original data back for the wavelength range that covers the \ce{^{13}CO_2} band at 4.39 \ce{\mu}m. The reason for adding the original data back is to avoid introducing artificial features to the ice bands or smoothing spectral features as a result of the spline function. 

This procedure gives nearly identical spectra as that adopted by \citet{rubinstein2023}. The line-subtracted spectra are shown in the second column of Figure \ref{fig:overview}. Finally, the observed fluxes \textit{\ce{F^{obs}_{\lambda}}} were converted to optical depth scale by fitting a local continuum with a third-order polynomial and using equation (\ref{eq:OD}),

\begin{equation}
\label{eq:OD}
    \tau^{obs}_{\lambda} = -ln\biggl(\frac{F^{obs}_{\lambda}}{F^{cont}_{\lambda}}\biggl),
\end{equation}
where \textit{\ce{F^{cont}_{\lambda}}} is the flux of the continuum. The \ce{^{13}CO_2} ice bands are shown in optical depth scale in the third column of Figure \ref{fig:overview}. 

It is worth noting that the 4.39 \ce{\mu}m asymmetric stretching mode of \ce{^{13}CO_2} lies at the very edge of the CO gas-phase line forest (Figure \ref{fig:Hops370-isolated}). Therefore, the ice band is not as strongly affected by the rotational-vibrational lines as other absorption features in this spectral range (e.g., the \ce{OCN^-} and OCS bands at 4.60 and 4.90 \ce{\mu}m, respectively). One minor hindrance is the \ce{H_2} 0 - 0 S(10) emission line at 4.41 \ce{\mu}m (Figure \ref{fig:overview}) that complicates the fitting of the local continuum due its location on the red wing of the \ce{^{13}CO_2} band.

\subsection{Spectral decomposition}

A comprehensive analysis of the observed spectral features warrants a critical approach when decomposing the ice absorption bands in order to avoid degenerate solutions. In this study we analyze the 4.39 \ce{\mu}m asymmetric stretching mode of \ce{^{13}CO_2} using two complementary approaches. The first is a phenomenological approach where we consistently fit the absorption bands of all the sources using a minimum number of common individual Gaussian profiles. Each Gaussian profile is representative of \ce{CO_2} in a specific molecular environment. This approach allows us to find the simplest unbiased way of "modeling" this ice profile by looking at the specific spectral features comprising the band. For this method we varied the widths and the peak positions of the Gaussian curves and first fit the ice absorption feature of the source that has the most extreme ice band profile, in our case IRAS 20126. Once the peak position and full width at half maximum (FWHM) of these Gaussian curves were determined, we used the results to consistently fit the \ce{^{13}CO_2} ice profiles of the remaining sources with these same Gaussian components. This approach was also used for the \ce{^{13}CO_2} ice band analyses of the high-mass protostars in the ISO sample \citep{boogert1999iso}, for the decomposition of solid carbon monoxide (CO) ice profiles \citep{pontoppidan2003m} and to study the 15.2 \ce{\mu}m band of \ce{^{12}CO_2} ice \citep{pontoppidan2008c2d}. 

Following the phenomenological approach, we applied a second method and fitted the ice features with high resolution laboratory data \citep{ehrenfreund1997infrared, ehrenfreund1999laboratory, van2006infrared} taken from the Leiden Ice  Data Base for Astrochemistry \citep[LIDA;][]{rocha2022lida}. The combination of laboratory spectra selected for this analysis is based on the five-component decomposition of the 15.2 \ce{\mu}m \ce{^{12}CO_2} band \citep{pontoppidan2008c2d}. We then compared the results of the phenomenological approach and the laboratory analysis to determine if these same five components are also necessary to fit the weaker \ce{^{13}CO_2} band or if the solutions of the laboratory analysis are degenerate.

\begin{figure*}[h!]
    \centering
     \includegraphics[width=1\hsize]{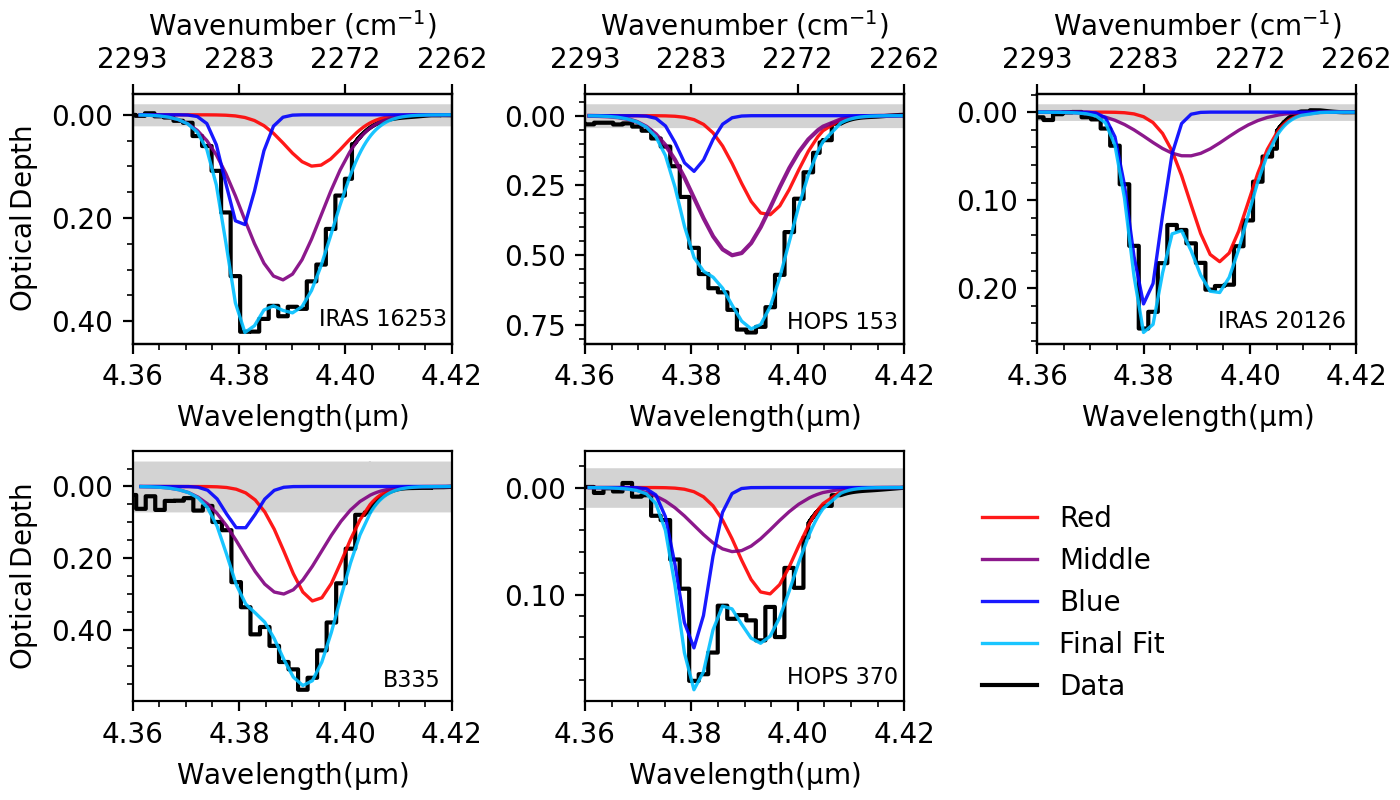}
    \caption{Spectral decomposition of the \ce{^{13}CO_2} bands using Gaussian profiles. The final fit is a linear combination of the three Gaussian curves, as indicated by the light blue line. The shaded area shows the rms in optical depth scale. After identifying the three components, the widths and central positions of the profiles are fixed for all five sources, leaving the relative optical depths of the components of the individual components as the only variables.}
    \label{fig:gaussians}
\end{figure*}

\section{Analysis}
\label{sec:3}

In this section we analyze the ice absorption bands of \ce{^{13}CO_2} for the five protostars in our sample. We first adopted a phenomenological approach to determine the minimum number of components required to fit the ice features \citep{boogert1999iso,pontoppidan2003m,pontoppidan2008c2d}. We then followed the methods described in \citet{pontoppidan2008c2d} and decomposed the bands using laboratory data.

\subsection{Spectral decomposition using a phenomenological approach}

\label{subsec:31}

We modeled the ice profiles of the \ce{^{13}CO_2} bands using linear combinations of Gaussian profiles that vary in width and central position. To determine the goodness of the fit, we used a routine that minimizes the root mean squared error (RMSE) between the final fit and the observed data. The RMSE was calculated using Equation \ref{eq:rmse}:

\begin{equation}
\label{eq:rmse}
    RMSE = \sqrt{\sum_{i=1}^{n} \frac{1}{n} \Biggl( \tau^{obs}_i - \sum_{j=1}^{m} \tau^{gauss}_j \Biggl) ^2 },
\end{equation}
where \ce{\tau^{obs}_i} is the observed data and \ce{\tau^{gauss}_j} is a Gaussian component. We obtained lower RMSE values for all the sources except IRAS 16253 when the ice features were modeled with three components instead of two thus indicating that a minimum of three Gaussian curves is needed to consistently model the bands. The three components are crucial in particular for fitting the double peaked profiles of HOPS 370 and IRAS 20126, since these spectral features cannot be modeled with only two components. 

The final fit comprises a red component centered at 2275.72 \ce{cm^{-1}} (4.39 \ce{\mu}m) on the long-wavelength side of the band, a blue component centered at 2282.92 \ce{cm^{-1}} (4.38 \ce{\mu}m) on the short-wavelength side of the band and a middle component at 2278.90 \ce{cm^{-1}} (4.39 \ce{\mu}m). The final fits are shown in  Figure \ref{fig:gaussians} and the best-fit parameters are given in Table \ref{Tab:gaus}. The interpretation of these components is discussed in Section \ref{subsec:41}

\begin{center}
\begin{table}[hbt!]
\caption{Properties of the Gaussian profile fit.}
\small
\centering
\begin{tabular}{lccc}
\hline \hline
Component &  Peak Position (\ce{cm^{-1}})  & FWHM (\ce{cm^{-1}} ) 
\\     
\hline

Red  & 2275.72 \ce{\pm} 0.1 & 6.6 \ce{\pm} 0.1 \\
Middle & 2278.90 \ce{\pm} 0.4 & 9.0 \ce{\pm} 0.6 \\
Blue & 2282.91 \ce{\pm} 0.01 & 3.5 \ce{\pm} 0.1 \\

\hline
\end{tabular}
\label{Tab:gaus}
\begin{tablenotes}\footnotesize
\item{} 
\end{tablenotes}
\end{table}  
\end{center}

We note that although we obtain good overall fits for all the ice profiles, the red wings of ice bands where we removed the gas phase lines and fitted the continuum introduce a small uncertainty in our solutions. In the following section we analyze the bands using laboratory data. 

\begin{center}
\begin{table*}[hbt!]
\caption{Laboratory spectra.}
\small
\centering
\begin{tabular}{lcccl}
\hline \hline
Ice sample & Ratio & $T$(K) & Resolution (\ce{cm^{-1}}) & Reference   \\ 
         
\hline

\ce{CO_2}:\ce{H_2O} & 1:10 & 10 & 1 & \citet{ehrenfreund1999laboratory}  \\

\ce{CO_2}:\ce{H_2O} & 1:10 & 160 & 1 & \citet{ehrenfreund1999laboratory}  \\

\ce{CO_2}:\ce{CH_3OH}  & 1:10 & 10 & 1 & \citet{ehrenfreund1999laboratory} \\

\ce{CO_2}:\ce{CH_3OH} & 3:1 & 105 & 1 & \citet{ehrenfreund1999laboratory} \\

\ce{CO_2}:CO & 1:1 & 15 & 0.5 &  \citet{van2006infrared} \\

\ce{CO_2}:CO & 1:2 & 25 & 0.5 &  \citet{van2006infrared} \\

\ce{CO_2} & Pure & 80 & 1 & \citet{ehrenfreund1997infrared} \\

\hline
\end{tabular}
\label{Tab:2}
\begin{tablenotes}\footnotesize
\item{\textbf{Notes.} List of laboratory ice compositions used in this work. All corresponding spectra are available for public use on the Leiden Ice Data Base for Astrochemistry \citep{rocha2022lida}.} 
\end{tablenotes}
\end{table*}  
\end{center}

\begin{center}
\begin{table*}[hbt!]
\caption{Optical depths of individual components.}
\small
\centering
\begin{tabular}{lccccccc}
\hline \hline
Ice sample & Ratio & $T$ (K) & \ce{\tau_{IRAS 16253}} & \ce{\tau_{B335}} & \ce{\tau_{HOPS 153}} & \ce{\tau_{HOPS 370}} & \ce{\tau_{IRAS 20126}} \\ 
         
\hline

\ce{CO_2}:\ce{H_2O} & 1:10   & 10 & 0.31 & 0.41 & 0.62 & 0.10 & 0.09  \\

\ce{CO_2}:\ce{CH_3OH} & 1:10  & 10 & 0.06 & 0.22 & 0.26 & 0.07 & 0.14 \\

\ce{CO_2}:CO & 1:1 & 15 & 0.11 & 0.07 & 0.08 & 0.02 & 0.02  \\

\ce{CO_2}:CO & 1:2 & 25 & 0.10 & 0.06 & 0.10  & 0.07  & 0.04   \\

\ce{CO_2} & Pure & 80 & 0.06 & 0.03 & 0.04 & 0.05  & 0.15 \\

\hline

\ce{CO_2}:\ce{H_2O} & 1:10 & 160 & - & -  & - & 0.10 & 0.15 \\

\ce{CO_2}:\ce{CH_3OH} & 3:1  & 105  & - & - & - & 0.13 & 0.15 \\

\ce{CO_2} & Pure & 80 & - & - & - & 0.06  & 0.12 \\

\hline
\end{tabular}
\label{Tab:OD}
\begin{tablenotes}\footnotesize
\item{\textbf{Notes.} Optical depths of individual components contributing to the final linear combination as discussed in Section 3.2.6. The bottom three rows provide optical depths for the alternative analysis of HOPS 370 and IRAS 20126} 
\end{tablenotes}
\end{table*}  
\end{center}

\subsection{Spectral decomposition based on the 15.2 \ce{\mu}m band}

\label{subsec:32}

To analyze the ice features with laboratory spectra, we followed the steps described in \citet{pontoppidan2008c2d} for the bending mode of \ce{^{12}CO_2} at 15.2 \ce{\mu}m. It was shown that this band can be decomposed into five unique components, each attributed to \ce{CO_2} embedded in different ice matrices. The band therefore acts as a powerful tracer of the chemical and thermal environment of the ice. The first component was assigned to \ce{CO_2} trapped in a water-rich ice matrix, and it has a contribution predominantly on the red side of the  \ce{^{12}CO_2} band. The second component, which was observed in several sources as a "red shoulder," was attributed to a mixture of \ce{CO_2} and \ce{CH_3OH} ice. The third component is a narrow feature on the blue side of the 15.2 \ce{\mu}m band and corresponds to an ice matrix where \ce{CO_2} is diluted in pure CO. An additional broad component corresponding to \ce{CO_2} mixed with \ce{CO} in almost equal parts also contributes to the blue-wing of the {15.2 \ce{\mu}m band. This component was described as a "variable" component since the exact ratio of \ce{CO_2} and CO can vary depending on the source. The final component was assigned to pure \ce{CO_2} ice, which produces a distinctive double peak feature seen in multiple sources. 

Using these components as guidelines, we fitted the \ce{^{13}CO_2} bands for four out of the five IPA sources. Additionally, we expanded our study by including laboratory spectra with different mixing ratios and different temperatures in our analysis since these parameters can significantly impact the width and peak position of the ice bands. For the majority of the components, we used low temperature spectra taken at 10 K and 15 K with the exceptions of \ce{CO_2}:CO 1:2 at 25 K, pure \ce{CO_2} at 80 K, \ce{CO_2}:\ce{CH_3OH} 3:1 at 105 K and \ce{CO_2}:\ce{H_2O} 1:10 at 160 K. This selection of spectra is summarized in Table \ref{Tab:OD} and Figure \ref{fig:BoogertFWHM} and further discussed in the following sections. While some of these temperatures may appear high, the values become more plausible once we take into account the fact that interstellar temperatures are not directly equivalent to the temperatures recorded in the laboratory but are actually significantly lower since interstellar physicochemical processes take place over longer timescales and at lower pressures. This implies that changes in the ice structure, such as crystallization, take place at much lower temperatures \citep{boogert1999iso, minissale2022thermal,litgernik2023overview}.  

Similar to the phenomenological approach, we used a routine that  minimizes the RMSE by varying the relative optical depths of the laboratory spectra. The spectral analyses for IRAS 16253, B335, HOPS 153, and IRAS 20126 are shown in Figure \ref{fig:labfits} and the optical depths for each component are given in Table \ref{Tab:OD}. The analysis for HOPS 370 is discussed in Section 3.2.6.

\begin{figure*}[h!]
    \centering
     \includegraphics[width=1\hsize]{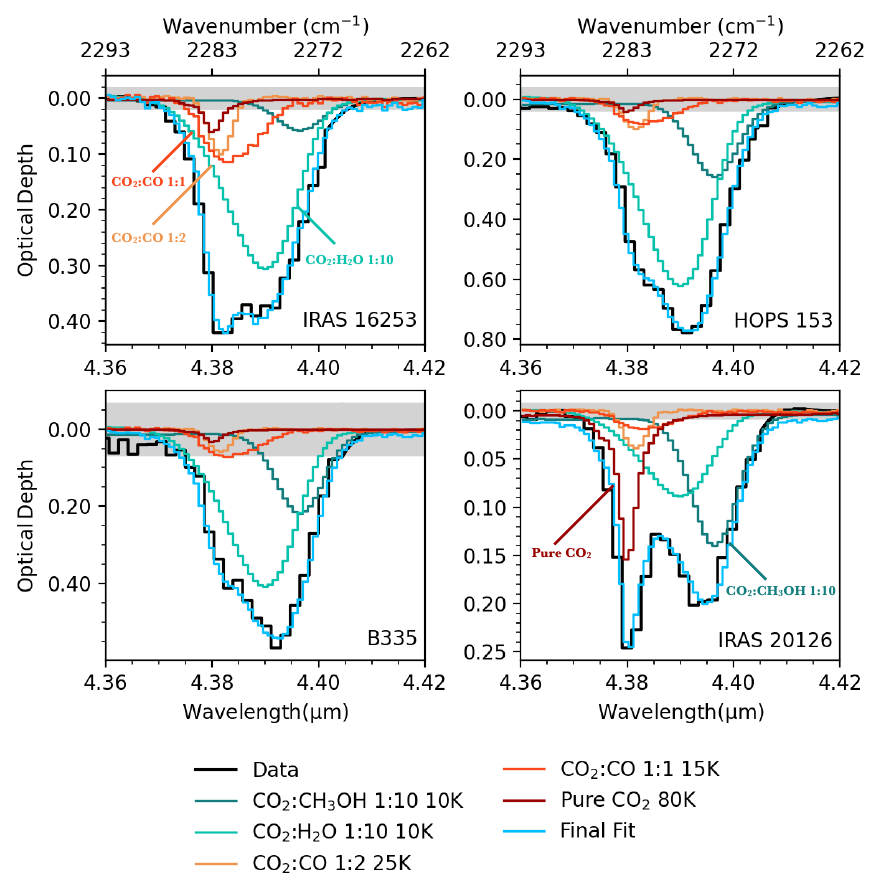}
    \caption{Decomposition of the \ce{^{13}CO_2} bands using selected laboratory spectra. The black line shows the observed spectrum, and the blue line shows the linear combination of all five different components. The dark green line corresponds to the \ce{CO_2}:\ce{CH_3OH} component. The aquamarine line shows the contribution of the broad \ce{CO_2}:\ce{H_2O} component. The gold line corresponds to the diluted \ce{CO_2}:\ce{CO} component, while the orange line shows the contribution of \ce{CO_2} and \ce{CO} mixed in equal parts. Finally, the dark red line corresponds to the pure \ce{CO_2} component. The shaded area shows the 3\ce{\sigma} rms in optical depth scale. The results for HOPS 370 are shown in a separate analysis.}
    \label{fig:labfits}
\end{figure*}

\begin{figure*}[h!]
    \centering
     \includegraphics[width=1\hsize]{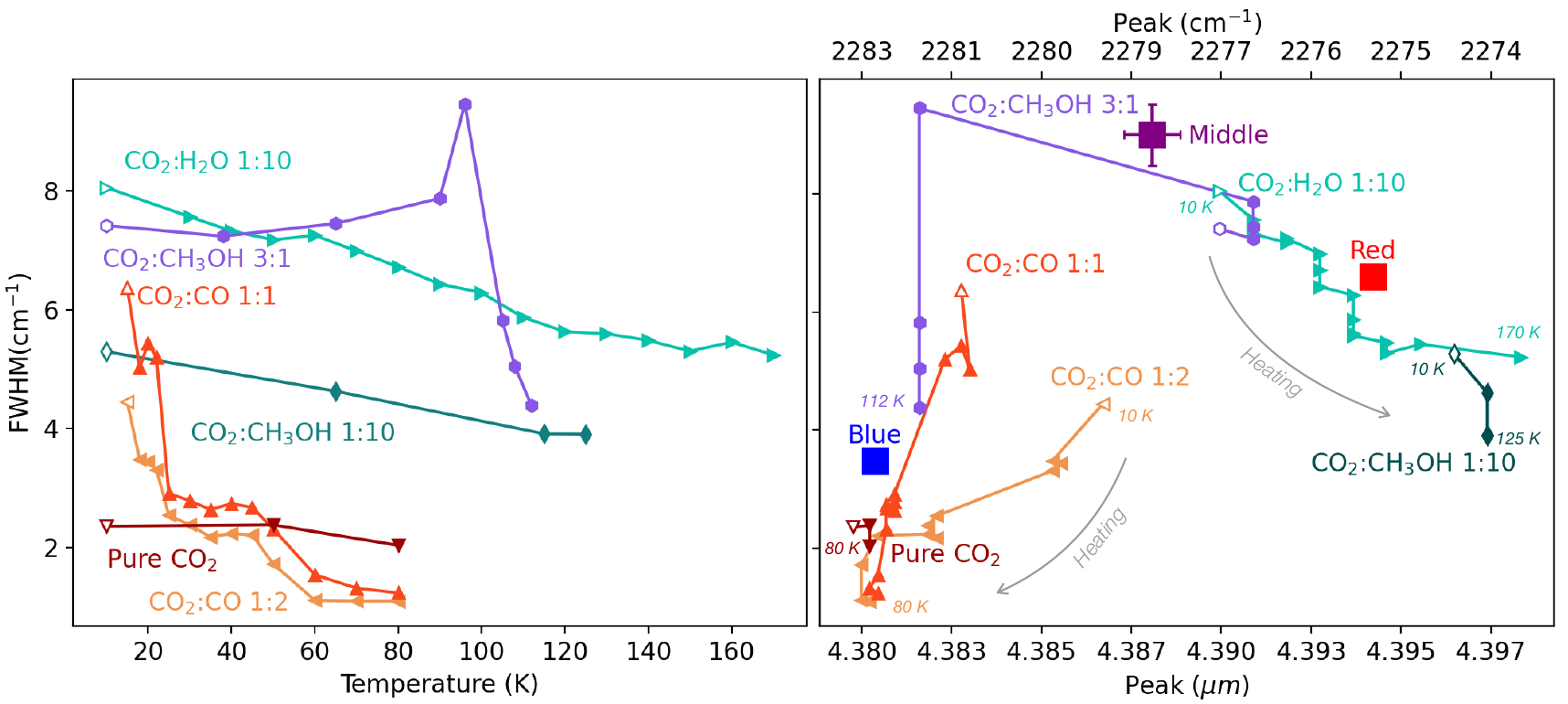}
    \caption{Properties of the laboratory spectra used in this work. Left panel: FWHM as a function of the temperature for the ice mixtures. The band profile narrows or broadens with increasing temperature depending on the ice sample. Right panel: FWHM plotted against the peak position. The peak centers of the \ce{CO_2}:\ce{H_2O} (aquamarine) and \ce{CO_2}:\ce{CH_3OH} (dark green) polar ices shift to longer wavelengths as the ice is heated. Similarly, the peak center of pure \ce{CO_2} ice also shifts to longer wavelengths with increasing temperature (dark red). The apolar \ce{CO_2}:CO (gold and orange) ices show the opposite behavior, with peaks shifting to shorter wavelengths as a function of temperature. The blue, red, and purple squares show the FWHM and peak positions of the fitted Gaussian profiles. } 
    \label{fig:BoogertFWHM}
\end{figure*}

\textit{\subsubsection{The \ce{H_2O}-rich broad component}}
\label{sec:3.1}

Similar to the \ce{^{12}CO_2} band at 15.2 \ce{\mu}m, the spectra of all the targets are dominated by the broad band of \ce{CO_2}:\ce{H_2O} 1:10 ice at 10 K (Figure \ref{fig:labfits}). This component was fitted with laboratory spectra taken by \citet{ehrenfreund1999laboratory}. The band is centered around 2277.93 \ce{cm^{-1}} (4.39 \ce{\mu}m), and it is the main contributor of the "red peak" of the \ce{^{13}CO_2} absorption feature. The peak center shifts significantly to longer wavelengths as the temperature increases from 10 K before desorbing at 175 K (Figure \ref{fig:BoogertFWHM} right panel). Additionally, the width of the band decreases with increasing temperature (Figure \ref{fig:BoogertFWHM} left panel). Of the available and tested spectra we found that the spectrum at 10 K provides the best fit, which is in agreement with the temperature suggested by \citet{pontoppidan2008c2d} to fit the \ce{^{12}CO_2} 15.2 \ce{\mu}m band. 
    
\textit{\subsubsection{The \ce{CH_3OH}-rich red component}}
\label{sec:3.2}

In addition to the water-rich component, we also included the \ce{CO_2}:\ce{CH_3OH} 1:10 laboratory spectrum \citep{ehrenfreund1999laboratory} in this analysis. At 15.2 \ce{\mu}m this component was observed as a "red shoulder" and for the \ce{^{13}CO_2} band this methanol-rich component is required to properly fit the red wing of the band (Figure \ref{fig:labfits}). The peak of this component is centered at 2274.55 \ce{cm^{-1}} (4.40 \ce{\mu}m) somewhat redshifted compared to the \ce{CO_2}:\ce{H_2O} 1:10 component that peaks at 2277.93 \ce{cm^{-1}} (4.39 \ce{\mu}m) at 10 K. The band shifts to slightly longer wavelengths as a function of temperature (Figure \ref{fig:BoogertFWHM} right panel). A narrowing of the band is also observed as the temperature of the ice increases (Figure \ref{fig:BoogertFWHM} left panel). After considering the candidate spectra, we found that the spectrum at 10 K provides the best fit. 

\textit{\subsubsection{The pure \ce{CO_2} component}}
\label{sec:3.3}
An additional peak on the blue side of the band is observed for the high luminosity sources HOPS 370 and IRAS 20126. The peak can be reproduced using laboratory spectrum of pure \ce{CO_2} at 80 K \citep{ehrenfreund1997infrared}. The spectrum has a narrow band with a peak centered around 2282.99 \ce{cm^{-1}} (4.38 \ce{\mu}m) that shifts to slightly longer wavelengths as a function of the temperature (Figure \ref{fig:BoogertFWHM} right panel). While this component is dominant in IRAS 20126 and HOPS 370, it also has a small contribution in B335 and HOPS 153 where the component is required to improve the fit of the blue wing of the bands. In IRAS 16253, the contribution of the component is significant enough that it hints at the blue peak. From the temperature analysis we deduced that the spectrum at 80 K provided the best overall fit due to its slightly redshifted central position.

\textit{\subsubsection{{The narrow blue component}}}
\label{sec:3.4}

An additional contribution of apolar \ce{CO_2}:CO 1:2 ice \citep{van2006infrared} was also included in the linear combination to better fit the blue peak. The peak position of this band is centered at 2282.02 \ce{cm^{-1}} (4.38 \ce{\mu}m), and shifts to shorter wavelengths when the ice is heated (Figure \ref{fig:BoogertFWHM}, right panel). The width of the peak also decreases significantly between 15 K and 25 K (Figure \ref{fig:BoogertFWHM}, left panel). We tested laboratory spectra at different temperatures and determined that the spectrum at 25 K provided the best fit because of its width and peak position. We note that although 25 K is close to the desorption temperature of CO, a fraction of the molecules can still be present in the ice as they remain trapped in the matrix of the less volatile \ce{CO_2} \citep{fayolle2011laboratory}. The contribution of this component is 
indispensable since its peak position is slightly redshifted compared to the pure \ce{CO_2} component, which peaks at 2282.99 \ce{cm^{-1}} at 10 K. As a result, this component in combination with the narrow pure \ce{CO_2} component reproduce the required width to properly fit the blue peak. This \ce{CO_2}:CO component is the equivalent of the "dilute" component suggested in \citet{pontoppidan2008c2d}. We also tested solutions where 
\ce{CO_2} is diluted in mixtures containing higher amounts of CO but a cross-check analysis revealed that solutions containing highly diluted \ce{CO_2}:CO mixtures will overproduce the \ce{^{12}CO} band at 4.67 \ce{\mu}m.

\textit{\subsubsection{{The broad blue component  }}}
\label{sec:3.5}
The final component in the analysis is the \ce{CO_2}:CO 1:1 spectrum from \citet{van2006infrared}. The profile of this band at 15 K is broader than that of the "diluted" \ce{CO_2}:CO 1:2 component, and its peak position is centered at 2281.66 \ce{cm^{-1}} (4.38 \ce{\mu}m), slightly redshifted compared to the diluted component at 25 K (Figure \ref{fig:BoogertFWHM} right panel). It is required to specifically fit the region between the red and the blue peak and is therefore essential in HOPS 370, IRAS 20126, and IRAS 16263. The peak center shifts to shorter wavelengths and narrows with increasing temperature. From the temperature analysis we determined that the spectrum at 15 K provided the best fit. This corresponds to the variable component suggested in \citet{pontoppidan2008c2d}, where \ce{CO_2} and CO are mixed in almost equal parts.

\textit{\subsubsection{{The analysis of HOPS 370}}}
\label{subsec:3.6}

\begin{figure*}[h!]
    \centering
     \includegraphics[width=\hsize]{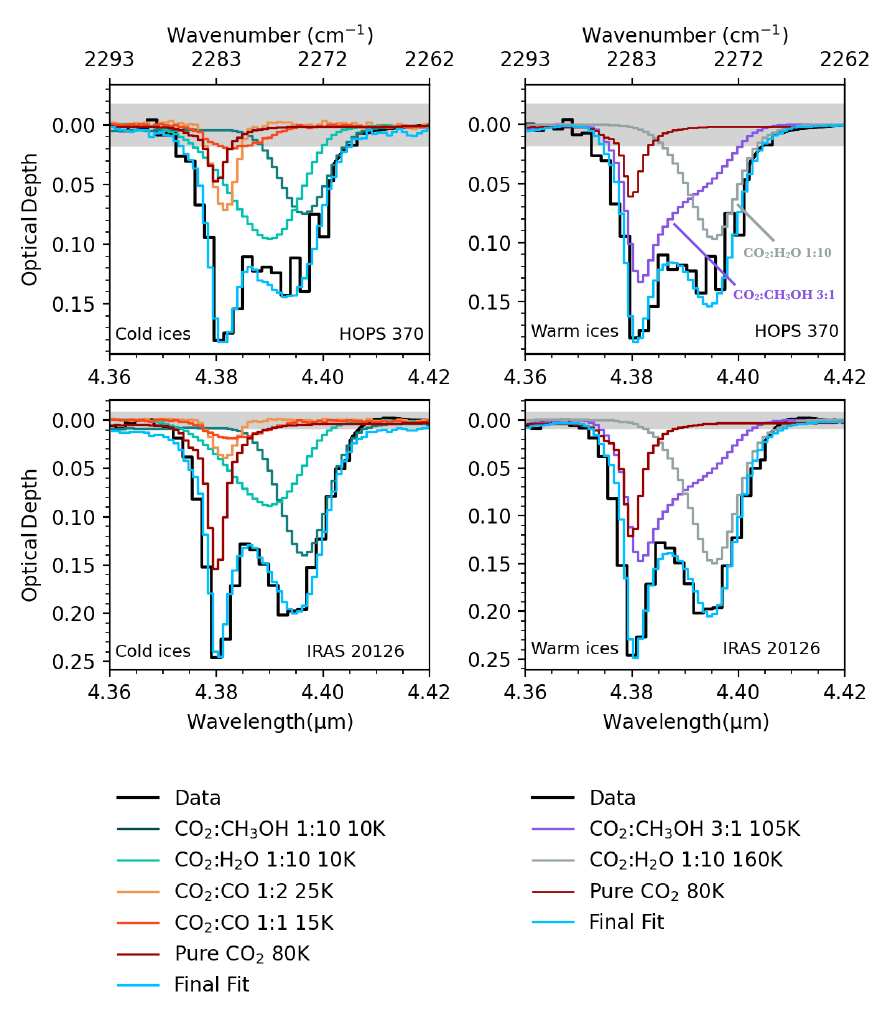}
    \caption{Alternative spectral analysis for HOPS 370 with high temperature ice species. The black line shows the observed spectrum, and the blue line shows the linear combination of all five different components. The purple line corresponds to the hot \ce{CO_2}:\ce{CH_3OH} component. The solid gray line shows the contribution of the hot \ce{CO_2}:\ce{H_2O} component. Finally, the dark red line corresponds to the pure \ce{CO_2} component. The shaded area shows the 3\ce{\sigma} rms in optical depth scale} 
    \label{fig:hotices}
\end{figure*}

While the methods discussed in the previous sections provide good overall fits for all the targets, further analysis shows that HOPS 370 is a special case. It is the second most luminous source in the sample (Table \ref{Tab:1}) and requires a significant contribution from the apolar ices to fit the blue peak. We discovered, however, that the \ce{CO_2}:CO 1:1 and 1:2 laboratory spectra used to fit this blue peak were significantly over-fitting the absorption band of \ce{^{12}CO} at 4.67 \ce{\mu}m, which we used to cross-check the solutions we found for the \ce{^{13}CO_2} bands. This is not the case for the other targets in the sample since they have stronger CO bands compared to HOPS 370. Therefore, in order to fit the \ce{^{13}CO_2} ice profile without overproducing the CO absorption band it was necessary to discard \ce{CO_2}:CO 1:1 and 1:2 components from the analysis and assume that \ce{CO_2}:CO ice is not (significantly) contributing to the \ce{^{13}CO_2} ice band of HOPS 370. 

Instead we fitted an alternative hot \ce{CO_2}:\ce{CH_3OH} 3:1 component \citep{ehrenfreund1999laboratory} to the blue peak. We opted for the spectrum at 105 K since at these higher temperatures the peak position of the \ce{^{13}CO_2} isotopologue band at 4.39 \ce{\mu}m is not only blueshifted enough to fit the blue peak (Figure \ref{fig:BoogertFWHM}), but at 15.2 \ce{\mu}m the band also produces both the shoulder and the double peak feature observed in the bending mode as illustrated in the laboratory results of \citet{ehrenfreund1999laboratory}. Additionally, we added the spectrum of hot water ice at 160 K to fit the red peak of the band thus fully replacing the cold \ce{CO_2}:\ce{CH_3OH} 1:10 component. Taking these alternative components in combination with the spectrum of pure \ce{CO_2} at 80 K we found a good fit for the \ce{^{13}CO_2} band without overproducing the CO ice band (Figure \ref{fig:hotices}). We were also able to fit the ice feature of IRAS 20126 using this same alternative analysis as illustrated in the second row of Figure \ref{fig:hotices}.

\hfill

\section{Discussion}

\label{sec:4}
In Section \ref{sec:3} we modeled the ice profiles of \ce{^{13}CO_2} using both a phenomenological approach where we fitted the bands with Gaussian profiles, and a mix-and-match method where we used high resolution laboratory data. The results from the phenomenological approach indicate that three components are sufficient to model the profile of the \ce{^{13}CO_2} ice band. Alternatively, the laboratory analysis shows that the same five-component decomposition used for the \ce{^{12}CO_2} ice absorption feature \citep{pontoppidan2008c2d}, can also be applied to the weaker isotopologue band. The spectra that we selected to model the ice profiles can be categorized into \ce{H_2O} - and \ce{CH_3OH}-rich polar ices, and apolar ices where \ce{CO_2} is embedded in a CO-rich ice matrix or present in pure crystalline form, consistent with previous studies of \ce{^{12}CO_2} and \ce{^{13}CO_2} ice bands \citep{pontoppidan2008c2d, boogert1999iso}.

In the following section we compare the results of these two different methods and show that, although the phenomenological approach provides the simpler solution, the physical laboratory data indicate that three components are not sufficient to model the ice band.\ This further strengthens  the argument that the 15.2 \ce{\mu}m decomposition method is also valid for this weaker isotopologue band.

\subsection{Comparison of the two approaches}

\label{subsec:41}
We compared the two approaches by looking at the spectral properties of the components used in both analyses, namely the FWHM and central positions of the Gaussian profiles and of the laboratory spectra. Looking first at the short wavelength side of the band, the width of the blue peak (Table \ref{Tab:gaus}) implies that more than two laboratory spectra are needed for that region of the band alone. While the entire peak was fitted with a single Gaussian curve, the profile of this curve is far wider than any of the laboratory spectra with central positions in this region as shown in Figure \ref{fig:BoogertFWHM} (right panel). The much narrower profiles of pure \ce{CO_2} ice bands are not sufficient to fully fit this blue peak, thus indicating that either a spectrum with a wider profile is needed or a second component is needed to contribute to the width. 

Similarly, the apolar \ce{CO_2}-CO ice profiles are also far too narrow to fit the peak, and the ice mixtures that do have a similar FWHM as the fitted Gaussian profile (e.g., the \ce{CO_2}:CO 1:2 mixtures, Figure \ref{fig:BoogertFWHM}) are too redshifted. An alternative ice band that could reproduce this width is the \ce{CO_2}:\ce{CH_3OH} 3:1 mixture (Figures \ref{fig:hotices} and \ref{fig:BoogertFWHM}) but this band is also slightly redshifted, thus still requiring the pure \ce{CO_2} component to properly fit the peak. 

The properties of the middle Gaussian profile (Table \ref{Tab:gaus}) also point toward an oversimplification of the linear solution. This is because the peak position and FWHM of this profile also do not coincide with any available laboratory data (Figure \ref{fig:BoogertFWHM}). More specifically, the peak position of this middle component is very blueshifted compared to the broad water-rich laboratory spectra to which it is normally assigned. We stress again that the laboratory spectra chosen in this work and the other candidate spectra that are explored in this study are based on detailed analysis of observed \ce{CO_2} bands conducted in previous studies. 

Finally, the red Gaussian curve (Table \ref{Tab:gaus}) used to fit the long wavelength region of the band is also wider and more blueshifted compared to the FWHM and peak positions of the methanol-rich spectra that are associated with this component. While the FWHM and central position of this profile do bring it in the range of the warm water-rich spectra ($\sim$ 80 K) (Figure \ref{fig:BoogertFWHM}), the blueshifted central position and broad profile of this band remain an issue when this band is included in an analysis with the other components. 

Therefore, we argue that although the phenomenological approach provides a simpler three-component solution, the spectral characteristics of these components do not match those of any known laboratory spectra. In light of these results, we conclude that the 15.2 \ce{\mu}m \ce{^{12}CO_2} solution can be adopted to study the ice profile of weaker \ce{^{13}CO_2} ice band. 

In the following sections we take a closer look at the origin of these ices in order to understand these various components that collectively comprise the \ce{^{13}CO_2} ice band.

\subsection{Tracing the chemical composition of the ices}
\label{subsec:4.2}
Formation of ices begins with the accretion of atoms and small molecules from the gas phase onto cold dust grains. Models \citep{garrod2011formation}, experimental results \citep{oba2010experimental,ioppolo2011surface} and studies toward quiescent clouds, low- and high-mass protostars \citep{bergin2005spitzer, whittet2007abundance, oberg2011spitzer} all indicate that, under dark cloud conditions, \ce{CO_2} forms efficiently along with \ce{H_2O} through reaction \ref{eq:pathway1}:

\begin{equation}
    \ch{CO + OH -> HO-CO -> CO_2 + H } \label{eq:pathway1}
.\end{equation}

In this scenario, oxygen atoms are hydrogenated and converted into \ce{H_2O} and OH radicals. These OH radicals will then react with CO accreted from the gas phase to first form the HO-CO complex, which then dissociates resulting in \ce{CO_2} ice. The co-formation route of \ce{CO_2} with \ce{H_2O} and the high efficiency of these processes produce the first generation \ce{H_2O}:\ce{CO_2} ices in high abundances, thus explaining the dominant presence of the water component in the \ce{CO_2} bands. The strength of this component in the bending mode at 15.2 \ce{\mu}m facilitated its detection in early studies by \citet{d1989discovery} and \citet{graauw1996sws}. \citet{boogert1999iso}, and \citet{gerakines1999infrared} also found that the \ce{^{13}CO_2} band of most sources in the ISO sample could be fitted with Gaussian profiles that corresponded to \ce{CO_2} residing in polar ices with several sources requiring only the polar component for an accurate fit. In observational spectra these polar signatures appear as broad bands peaking at longer wavelengths (Figure \ref{fig:BoogertFWHM}).

The formation of the \ce{H_2O}:\ce{CO_2} ice layers remains efficient at low extinction when the density of the cloud is not sufficiently high to enable catastrophic \ce{CO} freeze-out \citep{pontoppidan2006spatial}. Once the cloud crosses this threshold however, \ce{CO} is rapidly accreted onto the dust grains, forming a CO-rich ice layer. Any OH radical that freezes onto this CO layer will react and be converted into \ce{CO_2}, giving rise to the apolar \ce{CO}:\ce{CO_2} ices. The bands of these apolar ices are characteristically narrow and they peak at slightly shorter wavelengths (Figure \ref{fig:BoogertFWHM}). Finally, \ce{CH_3OH} also forms synchronously with \ce{CO_2} through hydrogenation of \ce{CO} ice \citep{watanabe2003the, fuchs2009astrometry, whittet2011observational,ioppolo2011surface,walsh2014complex}. These co-formation routes could therefore explain the presence of the \ce{CO_2}:\ce{CH_3OH} ices observed in the protostars. Similar to the \ce{H_2O}-rich ices, ices containing large quantities of \ce{CH_3OH} have broad profiles centered at longer wavelengths (Figure \ref{fig:BoogertFWHM}).

This multilayer formation of interstellar ices is also supported by the model predictions of \citet{garrod2011formation}, \citet{taquet2013water} and \citet{taquet2014multilayer}. In their work, the authors show the gradual growth of the ice mantles as a function of time and accentuate the fractional abundance of the dominating species at each phase. The models consistently demonstrate that the main stages of interstellar ice formation are the \ce{H_2O}-dominated period where \ce{CO_2} successfully forms in tandem with water, and a later dark cloud period dominated by the formation of CO ice where conversion of CO to \ce{CO_2} and CO to its main products (e.g., \ce{CH_3OH}) also occur. 

These different ice formation routes are reflected back in the components that comprise the \ce{^{13}CO_2} ice feature thus, making this isotopologue band a sensitive probe of ice composition. 

\begin{figure*}
    \centering
     \includegraphics[width=\hsize]{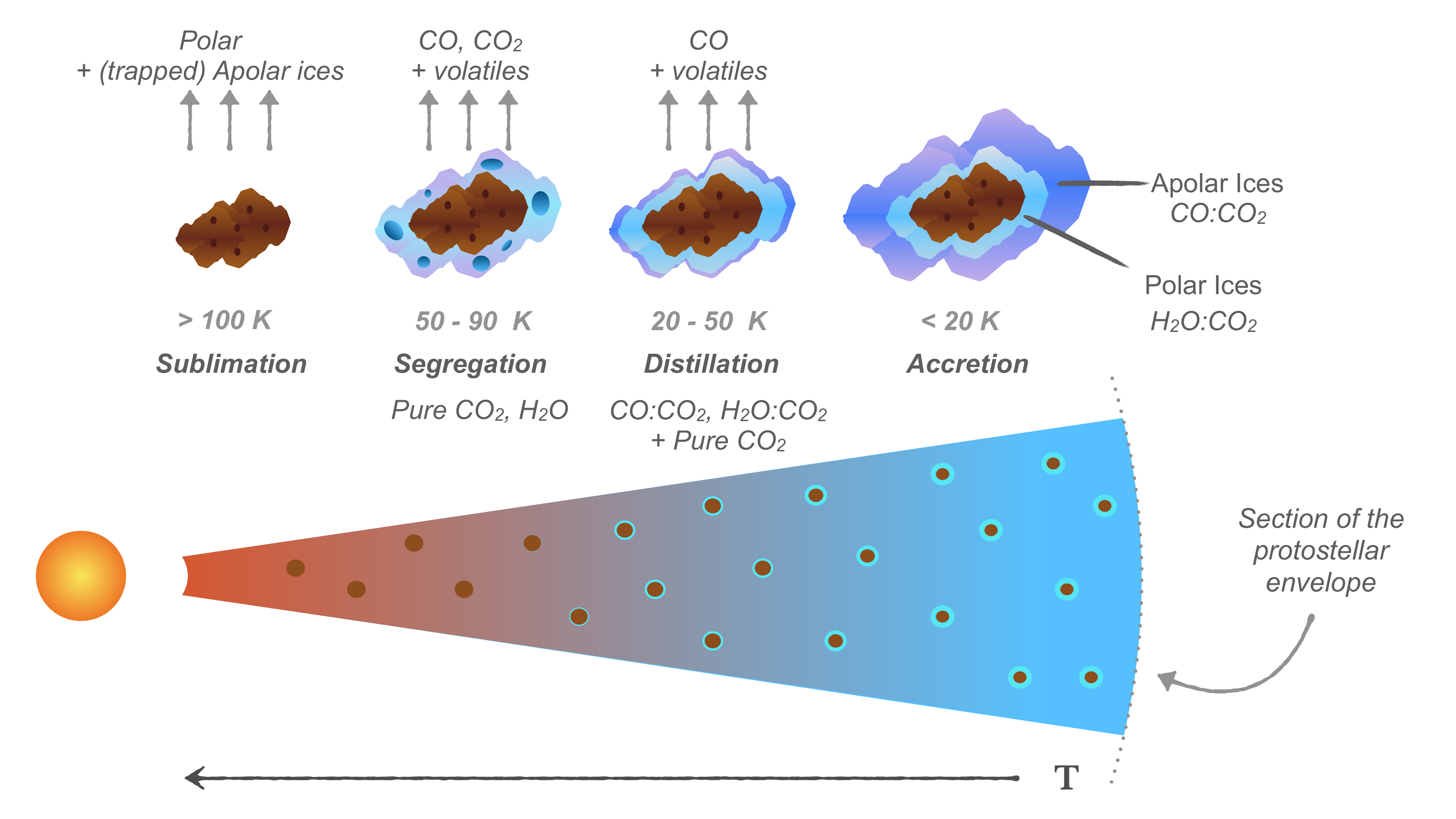}
    \caption{Schematic overview of ice processing mechanisms in protostars. The individual grains show the different stages of ice processing as a function of temperature. The dotted gray line traces the contour of the protostellar envelope. } 
    \label{fig:cartoon}
\end{figure*}

\subsection{Tracing the thermal history of the ices}
\label{subsec:4.3}
Following their formation, the overall abundance and structure  of interstellar ices can be impacted by thermal processing \citep{ehrenfreund1998ice,van1998chemical}. A\ rearranging of chemical bonds can cause distillation and segregation within the ice mantles \citep{pontoppidan2008c2d}, and as temperatures continue to rise the most volatile ices will desorb from the grains (Figure \ref{fig:cartoon}). The degree to which the ices are processed will depend first on the central source since high luminosity protostars will result in more ice heating. It is also determined by the location of the ices in the envelope as ice processing will increase with the increasing temperature gradient toward the central protostar (Figure \ref{fig:cartoon}). Additionally, intermittent heating periods in the context of episodic accretion will leave their mark in the ice inventory. There are indications, though, that the impact of such events is more critical for the CO ice budget compared to the \ce{CO_2} ice budget when the time interval to the previous outburst is much larger than the burst duration \citep{vorobyov2013the}. Finally, the evolutionary stage of the system also plays an important role since more evolved systems with lower envelope masses (high L/\ce{M_{env}}) tend to have elevated temperatures \citep{jorgensen2005molecular}. The observed ice bands are then the collective contribution of all the ice components on the individual grains and all the ices along the line of sight \citep{ehrenfreund1998ice}. This includes the processed ices located closer to the central protostar and the cold ices residing in the outer envelope, thus explaining the multiple components required to fit the spectral features. 

In Section \ref{sec:3} we show that, similar to the 15.2 \ce{\mu}m bending mode, the isotopologue band also shows significant variations that can be associated with ice processing. The bending mode is known to split into two discernible peaks, when pure \ce{CO_2} segregates from the other ices, due to ice heating \citep{sandford1990physical,ehrenfreund1997infrared, pontoppidan2008c2d, boogert2008c2d}. Among the five sources in our sample only IRAS 20126 and HOPS 370 display strong distinguishable blue peaks that match the laboratory spectra of pure \ce{CO_2}. These results are consistent with the ice segregation scenario since HOPS 370 and IRAS 20126 are the only high luminosity sources in the sample (310 and \ce{10^4} $L_\odot $, respectively), thus making it likely that larger volumes of ices in their envelopes have been subjected to protostellar heating. Our analysis in Section \ref{sec:3.2} also revealed that the \ce{^{13}CO_2} band of IRAS 16253 displays a hint of the additional blue peak that is associated with segregated pure \ce{CO_2} ice despite it being the lowest luminosity source in our sample. One reason why we are detecting these processed ices in this low-mass source could be the result of the accretion burst that it has recently experienced \citep{aso2023early, narang2023}.

We calculated the column density of the pure \ce{CO_2} component and determined its fraction with respect to the total \ce{^{13}CO_2} column density using equation \ref{eq:column}:

\begin{equation}
\label{eq:column}
    N = \frac{1}{A} \int \tau d\nu,
\end{equation}
where $A$ is the band strength corresponding to a specific ice absorption band and $\int$ $\tau$d$\nu$ is the integrated area under the ice absorption band.

For the band strengths of both the pure \ce{CO_2} component and the entire \ce{^{13}CO_2} absorption band we take the value derived for \ce{^{13}CO_2} in \citet{bouilloud2015}, A = 1.15 $\times$ $10^{-16}$ cm molecule$^{-1}$. For the uncertainties on the column densities, we used the uncertainties estimated for the band strengths, which are $\sim$ 20\%, and through error propagation we estimated the uncertainties on the derived ratios to be $\sim$ 30\%. The values of the pure \ce{CO_2} fraction for the five sources in our sample are shown in Figure \ref{fig:fraction}. The results indicate that HOPS 370 and IRAS 20126 have the largest fractions of pure \ce{CO_2} ice contributing to the blue peak, 9\% and 21\%, respectively. This is consistent with the fact that we expect higher levels of ice processing in the high luminosity sources. Of the remaining three lower luminosity sources, IRAS 16253 has the largest fraction of pure \ce{CO_2} ice (5 \%) and HOPS 153 has the smallest fraction (1.6 \%) despite having a higher luminosity than B335, which has a fraction of 2 \%. A recent study by \citet{kim2024} revealed that B335 also experienced a recent accretion burst, which could explain the increased fraction of pure \ce{CO_2} that we are observing.        

\begin{figure}
    \centering
     \includegraphics[width=\hsize]{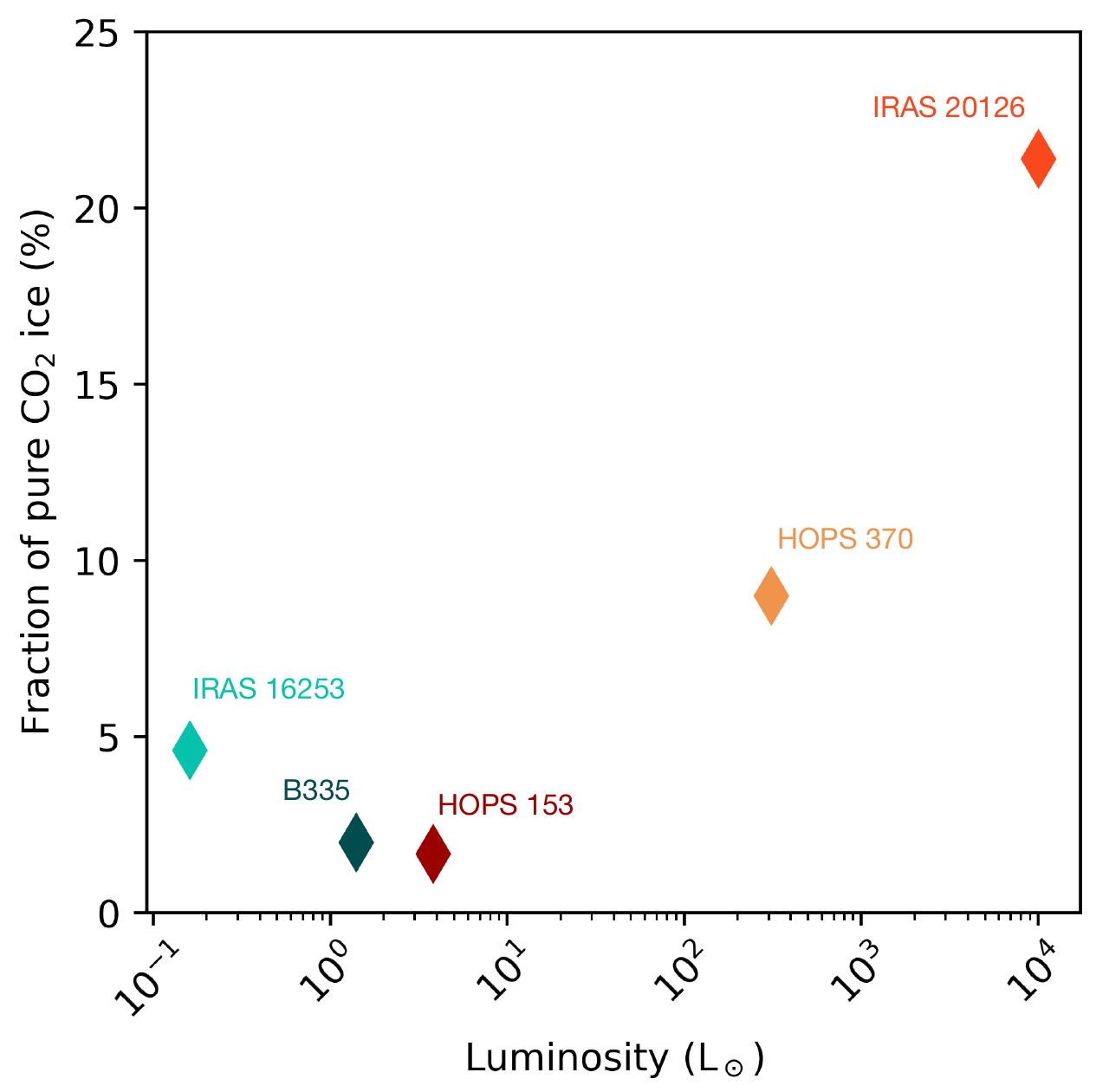}
    \caption{Fraction of pure \ce{CO_2} contributing to the total \ce{^{13}CO_2} ice band for the five different targets in this sample. The luminosity of each source is plotted in log-scale on the x-axis. } 
    \label{fig:fraction}
\end{figure}

The fact that we can observe significant levels of ice processing in the \ce{^{13}CO_2} ice band for the five different sources, indicates that this feature is good for diagnosing the thermal environment of these ices. This weaker isotopologue band becomes particularly useful in lines of sight where the \ce{^{12}CO_2} ice band at 15.2 \ce{\mu}m is saturated or is severely distorted by \text{grain shape and size} effects. It is therefore a useful alternative if not a better candidate to study ice morphologies.

\subsection{Thermal processing in HOPS 370}

Our analysis of HOPS 370 shows that a combination of warm and hot ices is needed in order to properly fit the asymmetric stretching mode of \ce{^{13}CO_2} at 4.39 \ce{\mu}m and that the cold CO:\ce{CO_2} component is fully excluded from the final solution. That an alternative solution is needed for this source is not entirely implausible given that \citet{poteet2013anomalous} also found that a different method involving high temperature ices was required to decompose the 15.2 \ce{\mu}m \ce{^{12}CO_2} absorption feature of HOPS 68. Additionally, since the main CO band at 4.67 \ce{\mu}m was detected at such low optical depths, we could expect a negligible contribution of the \ce{CO_2}:CO component at 4.39 \ce{\mu}m, thus explaining the decision to omit this component from the linear fit. We also successfully fit the \ce{^{13}CO_2} absorption band of IRAS 20126 (L = \ce{10^4}  $ L_\odot $) with this alternative solution. This further suggests that a linear fit consisting of warm and hot ices could be valid, and in some cases even necessary, for high-mass luminous sources.

We note that this alternative analysis implies that the \ce{^{13}CO_2} band of HOPS 370 consists of only warm and hot ices with essentially the same ingredients of \ce{H_2O} and \ce{CH_3OH} albeit in different mixing ratios. These high temperatures can be expected however once the other observed spectral features are taken into account. For instance, the spectrum of HOPS 370 displays strong ro-vibrational gas-phase lines (Figure \ref{fig:Hops370-hot}). These strong gas-phase lines in combination with the weak CO ice band indicate that the overall envelope temperature is high and that the CO ice might have sublimated (i.e., the coldest outer part of the envelope in Figure \ref{fig:cartoon} is missing). In addition to this, the peaked shape of the water absorption band observed at 3 \ce{\mu}m further suggests that we might be probing a hot environment. This is because this peaked profile is characteristic of crystalline water \citep{boogert2015observations} and cross-analysis on this band showed that it could only be fitted with laboratory spectra with $T$ > 140 K (Figure \ref{fig:Hops370-hot}). It is therefore safe to assume that HOPS 370 contains highly processed ices and that the \ce{CO_2}:{CO} component has a negligible contribution in this source due to the elevated temperatures throughout the envelope. The high luminosity of the central source ($L$ = 310  $ L_\odot $) could be the cause of this hot environment but we also note that HOPS 370 may have experienced a recent mass accretion burst that could have heated the ices in a relatively short period of time \citep{tobin2020vla}. Similarly, IRAS 20216 was recently classified as a variable source, most likely due to episodic accretion events \citep{massi2023thesoul}, which could again enhance the amount of ice heating in its envelope. That HOPS 370 has very little CO compared to IRAS 20126 despite having a lower luminosity could be due to different initial conditions of the clouds themselves, potential external heating \citep{jorgensen2005molecular} or internal heating due the recent accretion burst \citep{vorobyov2013the}, bringing the entire envelope above the desorption temperature of CO. 

We note that a solution containing heated ices was previously proposed in \citet{boogert1999iso}. The novelty in the alternative solution proposed in this work is that  while \citet{boogert1999iso} used heated ices to fit the short wavelength peak, the long wavelength peak was still fitted using cold ices whereas our solution consists of only hot and warm ices. The short wavelength peak is fitted with a hot \ce{CO_2}:\ce{H_2O} spectrum that is consistent with the crystalline shape of the water ice band at 3 $\mu$m (Appendix \ref{Appendix:A}). 

\subsection{Future outlook}

Future studies will be focused on confirming the findings in this paper with the other ice bands in the spectra. A cross-analysis with the \ce{^{12}CO_2} band at 15.2 \ce{\mu}m in particular is necessary to further support the results in this work. This will be possible in combination with existing MIRI observations. Moreover, analysis on the \ce{^{12}CO_2} and \ce{^{12}CO} ice bands at 4.27 \ce{\mu}m and 4.67 \ce{\mu}m, respectively will also be the topic of future work, in particular to study the effect of ice heating on these bands for HOPS 370 and IRAS 20126.   Additionally, the column densities for the individual components will be determined once the cross-check in the MIRI range is completed. Future higher resolution NIRSpec data will also enable us to study the blue peak in more detail given that the width of this pure \ce{CO_2} component is 0.004 \ce{\mu}m and will therefore require a resolving power of at least twice R = 1200 to fully resolve the narrow feature without having significant broadening effects due to convolution, a feat that will be possible with future NIRSpec observations using the G395H mode (R = $\lambda$ /$\Delta$ $\lambda$ = 2700). These higher resolution NIRSpec observation might also reveal subtle differences in the spectral profiles that might point more conclusively toward either the cold or the hot solution.  Finally, a quantitative analysis where the \ce{^{12}C}/\ce{^{13}C} ratios are determined and compared across the different sources will also be the focus of future work.

\section{Conclusions}
\label{sec:5}

We have analyzed JWST NIRSpec data of the 4.39 \ce{\mu}m asymmetric stretching mode of \ce{^{13}CO_2} ice for five protostars, covering a large range in mass and luminosity and fitted the absorption features using a phenomenological approach and, subsequently, laboratory data. The bands show rich variations in spectral appearance, which reflect the thermal and chemical history of the ices. 

\begin{itemize}
    \item The \ce{^{13}CO_2} bands can be decomposed into five unique components using laboratory spectra with different ice mixtures and temperatures. The results show that, similar to the \ce{^{12}CO_2} bending mode at 15.2 \ce{\mu}m, the isotopologue band comprises spectral contributions from  cold \ce{H_2O}-rich and \ce{CH_3OH}-rich polar ices, cold CO-rich apolar ices, and warm pure \ce{CO_2} ice at 80 K.
    \item For the more luminous sources HOPS 370 and IRAS 20126, an additional narrow peak begins to dominate the blue side of the bands, which is indicative of pure \ce{CO_2} ice. This pure \ce{CO_2} ice is likely the result of ice distillation and segregation due to heating of the ice. We propose an alternative analysis consisting of warm \ce{CO_2}:\ce{CH_3OH} and \ce{CO_2}:\ce{H_2O} ices and warm pure \ce{CO_2} ice at 80 K to fit the absorption bands of these two sources (i.e., no cold \ce{CO_2} ice component). The viability of this alternative analysis is supported by additional features in the spectra that point toward higher temperatures throughout the envelope.
    \item We confirm the validity of this multicomponent fit by demonstrating that, while a minimum of three Gaussian profiles does provide a good fit, the properties of these Gaussian profiles do not perfectly match any known laboratory data.

\end{itemize}

 With recent JWST data in the MIRI range (4.9 - 28 \ce{\mu}m), we can support our findings via a cross-analysis of the 15.2 \ce{\mu}m bands of these sources, determine the column densities, and constrain the \ce{^{12}C}/\ce{^{13}C} ratios.

\begin{acknowledgements}

Astrochemistry in Leiden is supported by the Netherlands Research School for Astronomy (NOVA), by funding from the European Re- search Council (ERC) under the European Union’s Horizon 2020 research and innovation programme (grant agreement No. 101019751 MOLDISK), and by the Dutch Research Council (NWO) grant 618.000.001. Support by the Danish National Research Foundation through the Center of Excellence “InterCat” (Grant agreement no.: DNRF150) is also acknowledged.

This work is based on observations made with the NASA/ESA/CSA\textit{ James Webb }Space Telescope. The data were obtained from the Mikulski Archive for Space Telescopes at the Space Telescope Science Institute, which is operated by the Association of Universities for Research in Astronomy, Inc., under NASA contract NAS 5-03127 for JWST.  These observations are associated with program \#1802. All the JWST data used in this paper can be found in MAST: \hyperlink{10.17909/3kky-t040}{http://dx.doi.org/10.17909/3kky-t040}. Support for SF, AER, STM, RG, WF, JG, JJT and DW in program \#1802 was provided by NASA through a grant from the Space Telescope Science Institute, which is operated by the Association of Universities for Research in Astronomy, Inc., under NASA contract NAS 5-03127.

ACG acknowledges from PRIN-MUR 2022 20228JPA3A “The path to star and planet formation in the JWST era (PATH)” and by INAF-GoG 2022 “NIR-dark Accretion Outbursts in Massive Young stellar objects (NAOMY)” and Large Grant INAF 2022 “YSOs Outflows, Disks and Accretion: towards a global framework for the evolution of planet forming systems (YODA)”.
NJE thanks the University of Texas at Austin for research support.
AS gratefully acknowledges support by the Fondecyt Regular (project
code 1220610), and ANID BASAL project FB210003. G.A. and M.O. acknowledge financial support from grants PID2020-114461GB-I00 and CEX2021-001131- S, funded by MCIN/AEI/10.13039/501100011033.

\end{acknowledgements}

\bibliographystyle{aa}
\bibliography{main} 

\begin{appendix}


\section{Additional figures and tables}
\label{Appendix:A}

\begin{center}
\begin{table}[hbt!]
\caption{Coordinates of extracted spectra.}
\small
\centering
\begin{tabular}{lcc}
\hline \hline
Source &  RA &Dec 
\\     
\hline

IRAS 16253 & 16:28:21.63 & -24:36:24.11 \\
B335 & 19:37:0.93 & 07:34:09.32 \\
HOPS 153 & 05:37:57.03 & -07:06:56.16 \\
HOPS 370 & 05:35:27.64 & -05:09:33.94 \\

IRAS 20126 & 20:14:26.04 & 41:13:32.43 \\

\hline
\end{tabular}
\label{Tab:coordinates}
\begin{tablenotes}\footnotesize
\item{} 
\end{tablenotes}
\end{table}  
\end{center}

\begin{figure*}[h!]
    \centering
     \includegraphics[width=0.7\hsize]{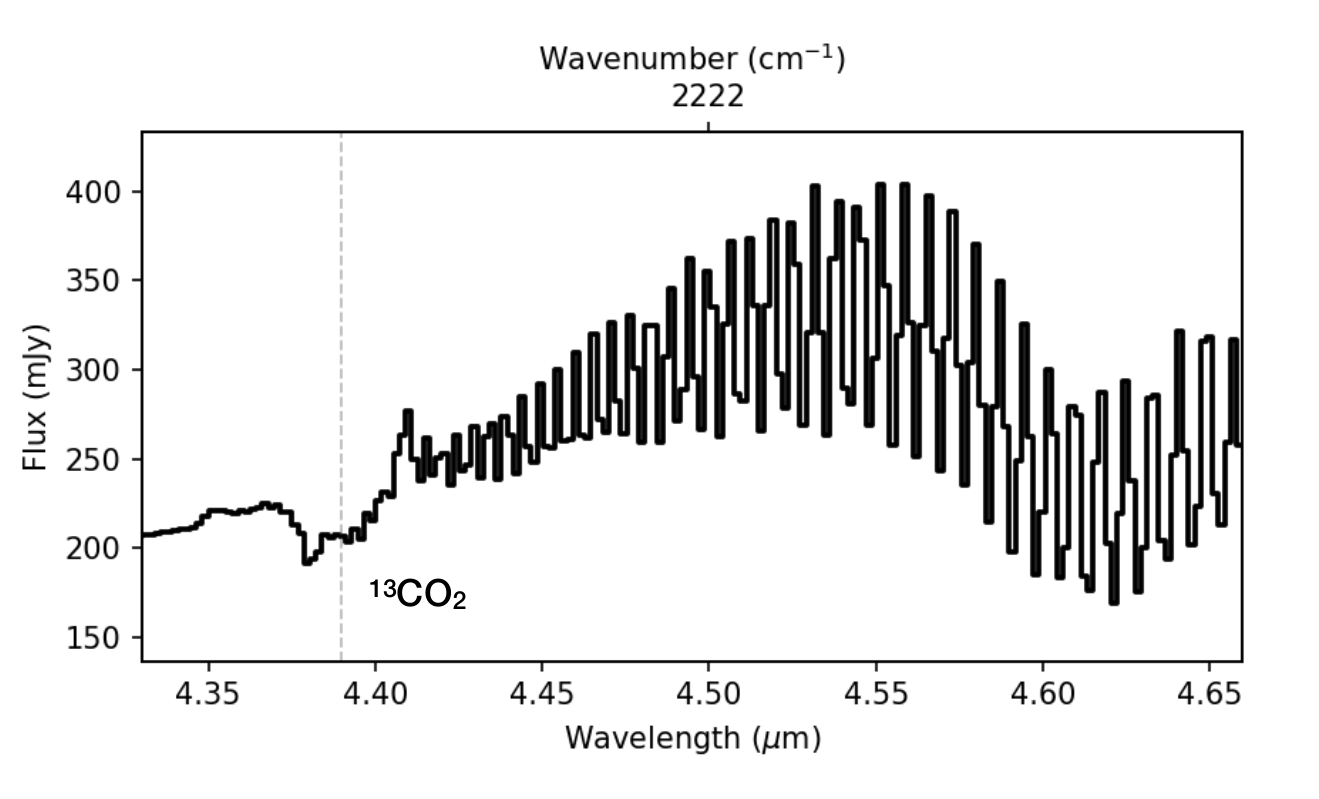}
    \caption{Spectrum of HOPS 370, the source that has the richest \ce{^{12}CO_2} rotational-vibrational lines \citep{federman2023,rubinstein2023}. Note that the \ce{^{13}CO_2} band lies on the edge of the CO line forest.} 
    \label{fig:Hops370-isolated}
\end{figure*}

\begin{figure*}[h!]
    \centering
     \includegraphics[width=1\hsize]{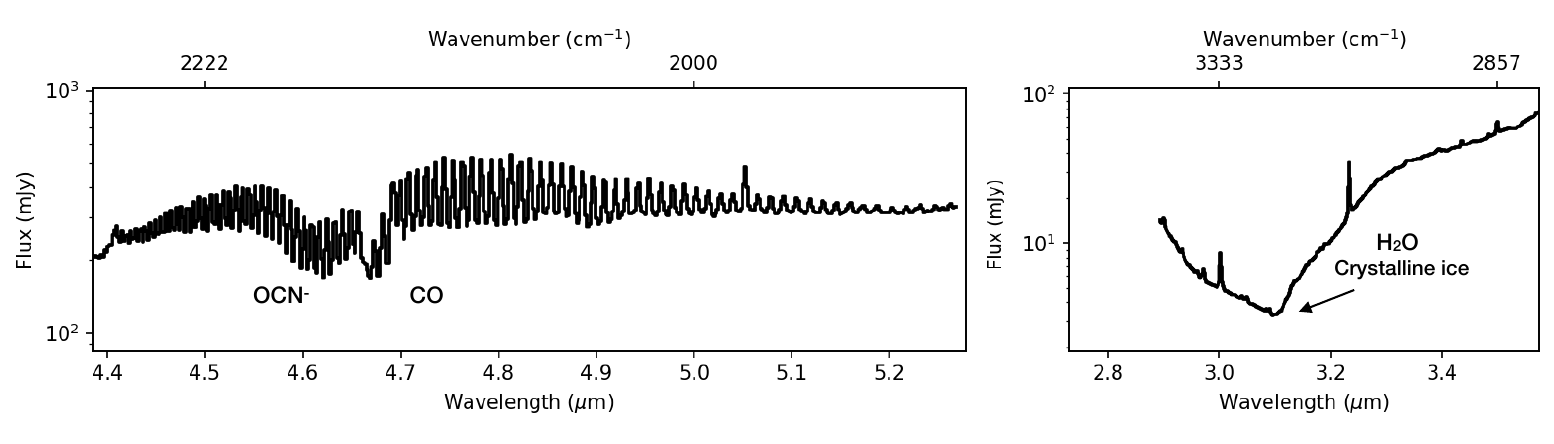}
    \caption{Hot spectral features toward HOPS 370. Left Panel:  4.67 \ce{\mu}m band of CO ice and the rotational-vibrational gas-phase lines. Note also the prominent \ce{OCN^-} ice feature at 4.60 \ce{\mu}m. Right panel:  3 \ce{\mu}m band of water. The peaked profile is characteristic of crystalline water.} 
    \label{fig:Hops370-hot}
\end{figure*}

\end{appendix}

\end{document}